\RequirePackage{ifpdf}
\ifpdf 
\documentclass[pdftex]{sigma}
\else
\documentclass{sigma}
\fi

\begin{document}
\allowdisplaybreaks

\renewcommand{\PaperNumber}{040}

\FirstPageHeading

\renewcommand{\thefootnote}{$\star$}

\ShortArticleName{Quantum Dynamics on the Worldvolume from Classical $su(n)$ Cohomology}

\ArticleName{Quantum Dynamics on the Worldvolume\\ from Classical $\boldsymbol{su(n)}$ Cohomology\footnote{This paper is a
contribution to the Special Issue on Deformation Quantization. The
full collection is available at
\href{http://www.emis.de/journals/SIGMA/Deformation_Quantization.html}{http://www.emis.de/journals/SIGMA/Deformation\_{}Quantization.html}}}

\Author{Jos\'e M. ISIDRO and Pedro FERN\'ANDEZ DE C\'ORDOBA}

\AuthorNameForHeading{J.M. Isidro and P. Fern\'andez de C\'ordoba}

\Address{Grupo de Modelizaci\'on Interdisciplinar Intertech,
Instituto Universitario de Matem\'atica Pura\\ y Aplicada,
Universidad Polit\'ecnica de Valencia,
Valencia 46022, Spain}
\Email{\href{mailto:joissan@mat.upv.es}{joissan@mat.upv.es}, \href{mailto:pfernandez@mat.upv.es}{pfernandez@mat.upv.es}}

\ArticleDates{Received February 05, 2008, in f\/inal form March
28, 2008; Published online April 15, 2008}

\Abstract{A key symmetry of classical $p$-branes is invariance under worldvolume dif\/feo\-morphisms. Under the assumption that the worldvolume, at f\/ixed values of the time, is a~compact, quantisable K\"ahler manifold,  we prove that the Lie algebra of volume-preserving dif\/feomorphisms of the worldvolume can be approximated  by $su(n)$, for $n\to\infty$. We also prove, under the same assumptions regarding the worldvolume at f\/ixed time,  that classical Nambu brackets on the worldvolume are quantised by the multibrackets corresponding to cocycles in the cohomology of the Lie algebra $su(n)$.}


\Classification{81T30; 17B56}

\renewcommand{\thefootnote}{\arabic{footnote}}
\setcounter{footnote}{0}

\section{Introduction}\label{za}

\subsection{Some history}

Ever since its inception by Wigner in 1932 \cite{WIGNER},  quantum mechanics on phase space has remained a source of inspiration for mathematical physicists trying to clarify the deep relation between the classical world and the quantum world.  Some milestones along this route were the pioneering articles~\cite{GROENEWOLD} and \cite{MOYAL}, which culminated in the formal development of deformation quantisation \cite{DQ}.  A central role in this programme is played by the $\star$-product operation on Poisson manifolds \cite{KONTSEVICH}.   Good references on the subject are \cite{CATTANEO, DITO}.

\subsection{Some mechanics}\label{zzchs}

Classical mechanics can be formulated on a Poisson manifold ${\cal M}$. The algebra ${\cal C}^{\infty}({\cal M})$ of smooth functions on ${\cal M}$ supports classical Poisson brackets (CPB), i.e., an antisymmetric, bilinear map
\begin{gather*}
\left\{\cdot\,,\cdot\right\}_{\rm CPB}\colon \ {\cal C}^{\infty}({\cal M})\times {\cal C}^{\infty}({\cal M})\longrightarrow {\cal C}^{\infty}({\cal M})
\end{gather*}
satisfying the Jacobi identity and the Leibniz derivation rule. Given a classical mechanics governed by a Hamiltonian function $H$, the time evolution of a time-independent function $f$ on phase space is determined by its CPB with $H$,
\begin{gather}
\frac{{\rm d}f}{{\rm d}t}=\left\{f, H\right\}_{\rm CPB}.
\label{zc}
\end{gather}
One can replace CPB with classical Nambu brackets (CNB) \cite{NAMBU}, in the following way. Assume that the system has $N$ degrees of freedom. The CNB of $2N$ functions $f_{1}, \ldots, f_{2N}$ is given by their Jacobian determinant with respect to the canonical coordinates $x^{i}$ and the conjugate momenta~$p^i$:
\begin{gather*}
\left\{f_1,\ldots,f_{2N}\right\}_{\rm CNB}={\rm alt}(\partial f_1,\ldots,\partial f_{2N})=\frac{\partial(f_1,\ldots,f_{2N})}{\partial(x^1,p^1,\ldots, x^N,p^N)}.
\end{gather*}
Then the CPB of any two phase-space functions $f,g$ can be expressed in terms of their CNB, with $N-1$ insertions of conjugate pairs $(x^i,p^i)$ and summing over the latter  \cite{CZ}:
\begin{gather}
\left\{f,g\right\}_{{\rm CPB}}=\frac{1}{(N-1)!}\left\{f,g,x^{i_1},p^{i_1},\ldots x^{i_{N-1}},p^{i_{N-1}}\right\}_{\rm CNB}.
\label{zd}
\end{gather}
A summation over all pairs of repeated indices is understood on the right-hand side of (\ref{zd}). This insertion of, plus summation over, conjugate pairs, is called {\it symplectic tracing} of the CNB. In this way, the equation of motion (\ref{zc}) becomes
\begin{gather}
\frac{{\rm d}f}{{\rm d}t}=\frac{1}{(N-1)!}\left\{f,H,x^{i_1},p^{i_1},\ldots,x^{i_{N-1}},p^{i_{N-1}}\right\}_{\rm CNB}.
\label{ze}
\end{gather}
Through the use of such symplectic traces, the equations of motion for an arbitrary classical system can be reexpressed
in terms of CNB. The possibility of so doing hinges on equation~(\ref{zd}). In turn, the latter rests on the CPB resolution of the CNB \cite{CZ}:
\begin{gather}
\left\{f_1,f_2,\ldots, f_{2N-1}, f_{2N}\right\}_{\rm CNB}
\nonumber\\
\qquad{} =\sum_{\pi\in P_{2N}}\frac{(-1)^{\pi}}{2^N N!}\left\{f_{\pi(1)},f_{\pi(2)}\right\}_{\rm CPB}\left\{f_{\pi(3)},f_{\pi(4)}\right\}_{\rm CPB}\cdots\left\{f_{\pi(2N-1)},f_{\pi(2N)}\right\}_{\rm CPB},
\label{zf}
\end{gather}
where $(-1)^{\pi}$ is the signature of the permutation $\pi$ within the symmetric group on $2N$ indices,~$P_{2N}$.

Upon quantisation, the algebra of functions $C^{\infty}({\cal M})$ on classical phase space gets replaced by the algebra of operators ${\cal O}({\cal H})$ on Hilbert space ${\cal H}$. The {\it quantum Nambu bracket} (QNB) of the operators $F_1,\ldots, F_{m}\in{\cal O}({\cal H})$ is def\/ined as the fully antisymmetrised product \cite{NAMBU}
\begin{gather}
[F_1,\ldots, F_{m}]_{\rm QNB}:=\mbox{alt}\,\left(F_1\cdots F_{m}\right)=\sum_{\pi\in P_{m}}(-1)^{\pi}\,F_{\pi(1)}\cdots F_{\pi(m)},
\label{zg}
\end{gather}
the usual quantum commutator being the particular case $m=2$. To avoid confusion with other brackets to be introduced later, we will use the notation \begin{gather*}
[F_1,F_2]_{\rm QPB}:= F_1F_2-F_2F_1, \qquad F_1,F_2\in {\cal O}({\cal H})
\end{gather*}
for the QNB of $m=2$ operators, the subindex QPB standing for {\it quantum Poisson brackets}. When $m=2k$, the following resolution of the QNB into a sum of products of QPB holds \cite{CZ}:
\begin{gather*}
[F_1,F_2,\ldots,F_{2k-1}, F_{2k}]_{\rm QNB}
=\sum_{\pi\in P_{2k}}\frac{(-1)^{\pi}}{2^kk!}\,[F_{\pi(1)}, F_{\pi(2)}]_{\rm QPB}\cdots [F_{\pi(2k-1)}, F_{\pi(2k)}]_{\rm QPB}.
\end{gather*}
This is the quantum analogue of the resolution (\ref{zf}), although it is valid for any $k$, not just $k=N$. The quantum analogue of equation~(\ref{zc}) is
\begin{gather*}
{\rm i}\hbar\,\frac{{\rm d}F}{{\rm d}t}=[F, H]_{\rm QPB},
\end{gather*}
while that of equation~(\ref{ze}), obtained also by symplectic tracing, reads
\begin{gather}
({\rm i}\hbar)^{N}\,\frac{{\rm d}F}{{\rm d}t}=\frac{1}{(N-1)!}\left[F,H,X^{i_1},P^{i_1},\ldots,X^{i_{N-1}},P^{i_{N-1}}\right]_{\rm QNB}.
\label{zk}
\end{gather}
Above, a factor of $({\rm i}\hbar)^{N-1}$ on the left-hand side accounts for $N-1$ symplectic traces on the right-hand side, while the remaining factor ${\rm i}\hbar$ is the one multiplying all time derivatives in the quantum theory.

\subsection{Some Lie-algebra cohomology}\label{rancio}

Along an apparently unrelated line, it turns out that multibracket operations on simple, compact  Lie algebras share many properties in common with Nambu brackets. These multibrackets can be constructed from a knowledge of the de Rham cohomology ring of the corresponding simple, compact Lie group  \cite{KNAPP}. For def\/initeness we concentrate on the group  $SU(n)$, whose Lie algebra $su(n)$ we denote, as usual, in lowercase. The Poincar\'e polynomial of $SU(n)$ is \cite{GOLDBERG}
\begin{gather*}
p_{SU(n)}(x)=(1+x^3)(1+x^5)\cdots (1+x^{2n-1}).
\end{gather*}
In plain words, $SU(n)$ contains cohomology cocycles in dimensions $3,5,\ldots,2n-1$, whose dimensions add up to $n^2-1$, the dimension of $SU(n)$.
At the Lie-algebra level this implies the existence of cohomology cocycles $\omega_3, \,\omega_5, \,\ldots,\, \omega_{2n-1}$. These cohomology cocycles of $su(n)$ are represented by totally antisymmetric tensors with as many indices as dictated by the corresponding dimensions. In order to compute their components let us consider the basis of $su(n)$ given by
\begin{gather}
[T_a, T_b]=\omega_{ab}^{\;\; c}\,T_c,
\label{zl}
\end{gather}
where the Killing metric reads, possibly up to an overall numerical factor,
\begin{gather}
k_{ab}={\rm tr}(T_aT_b).
\label{mazza}
\end{gather}
Then  the components $\omega_{abc}$ of the cocycle $\omega_3$ are given by
\begin{gather*}
\omega_{abc}=\omega_3(T_a,T_b,T_c).
\end{gather*}
The $\omega_{abc}$ can also be regarded as the result of lowering the upper index on the structure constants~$\omega_{ab}^{\;\; c}$, i.e.,
\begin{gather*}
\omega_{abc}=\omega_{ab}^{\;\; d}k_{dc}.
\end{gather*}
Consider now a $(2j+1)$-dimensional cocycle $\omega_{2j+1}$. In the basis (\ref{zl}), its components are given~by
\begin{gather}
\omega_{a_1a_2\ldots a_{2j+1}}=\omega_{2j+1}(T_{a_1},T_{a_2}, \ldots,T_{a_{2j+1}}).
\label{ilian}
\end{gather}
Using (\ref{ilian}) and  the inverse Killing metric $k^{ab}$ let us def\/ine
\begin{gather*}
\omega_{a_1\,a_2\,\ldots\,a_{2j}}^{\qquad\qquad b}:=\omega_{a_1\,a_2\,\ldots\,a_{2j},c}\,k^{cb}
\end{gather*}
On the $su(n)$ generators $T_a$ we def\/ine a $2j$-fold bracket by
\begin{gather}
[T_{a_1}, \ldots, T_{a_{2j}}]_{\omega_{2j+1}}=\omega_{a_1\ldots a_{2j}}^{\quad\quad b}\,T_b,
\label{zn}
\end{gather}
and extend it to all of $su(n)$ by multilinearity and complete antisymmetry. This ensures that the left-hand side of (\ref{zn}) is proportional to the determinant of the $2j\times 2j$ matrix whose rows are the $T_a$,
\begin{gather}
[T_{a_1} \ldots, T_{a_{2j}}]_{\omega_{2j+1}}=\lambda\,\mbox{alt}\,\left(T_{a_1} \cdots T_{a_{2j}}\right)
=\lambda\,\epsilon^{b_1\ldots b_{2j}}_{a_1\ldots a_{2j}}\,T_{b_1}\cdots T_{b_{2j}}.
\label{zm}
\end{gather}
The proportionality factor $\lambda$ can be chosen to be unity if the Killing metric (\ref{mazza}) is normalised appropriately. When $j=1$, (\ref{zn}) reduces to the Lie bracket (\ref{zl}). The operation (\ref{zm}) is identical to that of equation~(\ref{zg}), with the important dif\/ference, however, that the latter is an operation on quantum operators, while (\ref{zm}) is def\/ined on a classical Lie algebra. Moreover,  (\ref{zm}) requires an {\it even} number of entries, while the QNB of equation~(\ref{zg}) do not. One can express equation~(\ref{zm}) as a sum of products of 2-fold Lie brackets (\ref{zl}),
\begin{gather}
[T_{a_1},T_{a_2}\ldots,T_{a_{2j-1}},T_{a_{2j}}]_{\omega_{2j+1}}=\frac{1}{2^{j}}\,\epsilon^{b_1b_2\ldots b_{2j-1}b_{2j}}_{a_1a_2\ldots a_{2j-1} a_{2j}}\,[T_{b_1},T_{b_2}]\cdots [T_{b_{2j-1}},T_{b_{2j}}],
\label{zr}
\end{gather}
where the subindex $\omega_3$ is conventionally omitted from the binary brackets on the right-hand side. The above is analogous to the CPB resolution of the CNB of equation~(\ref{zf}), with $2j$-fold brackets replacing the CNB on the left-hand side, and 2-fold Lie brackets replacing the CPB on the right-hand side.

\subsection{Summary}\label{werty}

In this article we establish a relation between multibrackets on the Lie algebra of $su(n)$ and the quantum dynamics of $p$-branes, where $p=2k>2$ . We begin in Section~\ref{zs} by reviewing  a~similar link in the case of membranes. In Section~\ref{zzo} we move on to higher-dimensional, relativistic extended objects. We concentrate throughout on the bosonic sector of the corresponding $p$-bra\-nes. Our conclusions are presented in Section~\ref{ddkk}, where some examples are detailed and the relation with deformation quantisation is discussed.

\section{Membranes and Poisson brackets}\label{zs}

\subsection{Classical membranes}\label{zt}

The classical supermembrane in $\mathbb{R}^{11}$ is described by bosonic coordinates $x^{i}(\sigma_1,\sigma_2,\tau)$, $i=1, \ldots, 11$, and its superpartners. A key symmetry of the classical membrane is its invariance under {\it area-preserving reparametrisations} of the membrane worldvolume \cite{NICOLAI}. Inf\/initesimal, time-independent, area-preserving reparametrisations are coordinate transformations of the membrane worldvolume at f\/ixed time $\tau$, denoted $W_2$,
\begin{gather*}
\sigma^r\longrightarrow \sigma^r +\xi^r(\sigma), \qquad r=1,2,
\end{gather*}
such that
\begin{gather*}
\partial_r\left(w(\sigma)\xi^r(\sigma)\right)=0.
\end{gather*}
Above, $w(\sigma_1,\sigma_2)$ is a certain density on $W_2$ which we normalise to unity,
\begin{gather}
\int_{W_2}{\rm d}^2\sigma\,w(\sigma)=1.
\label{zw}
\end{gather}
Locally, all area-preserving reparametrisations can be expressed in terms of a single function~$\xi(\sigma)$, the dual on $W_2$ of the vector $\xi^r(\sigma)$,
\begin{gather}
\xi^r(\sigma)=\frac{\epsilon^{rs}}{w(\sigma)}\,\partial_s\xi(\sigma).
\label{zx}
\end{gather}
However, on topologically nontrivial worldvolumes, equation~(\ref{zx}) need not hold globally, so we will restrict to the subgroup generated by functions $\xi(\sigma)$ that are globally def\/ined on $W_2$. One can treat the Lie group $G_{W_2}$ of globally-def\/ined, area-preserving dif\/feomorphisms (an inf\/inite-dimensional group) in a way that resembles gauge theory. For this purpose one introduces the Lie brackets for functions on $W_2$
\begin{gather}
\big\{x^{i},x^{k}\big\}_{W_2}:= w(\sigma_1,\sigma_2)^{-1}\big(\partial_{\sigma_1}x^{i}\,\partial_{\sigma_2}x^k
-\partial_{\sigma_2}x^{i}\,\partial_{\sigma_1}x^k\big).
\label{zy}
\end{gather}
Then an inf\/initesimal area-preserving reparametrisation acts on the membrane's bosonic coordinates $x^{i}$ as
\begin{gather*}
\delta x^{i}=\left\{\xi, x^{i}\right\}_{W_2}.
\end{gather*}
A gauge f\/ield ${\cal A}$ taking values in the Lie algebra of $G_{W_2}$ can also be introduced, its transformation law being
\begin{gather*}
\delta{\cal A}=\partial_{\tau}\xi+\left\{\xi, {\cal A}\right\}_{W_2}.
\end{gather*}
Correspondingly we have the gauge covariant derivatives
\begin{gather*}
D_{\tau}x^{i}=\partial_{\tau}x^{i}+\left\{x^{i}, {\cal A}\right\}_{W_2}.
\end{gather*}
In the light-cone gauge only the transverse components $i=1,\ldots, 9$ are dynamical, and the bosonic part of the membrane Lagrangian reads \cite{NICOLAI}
\begin{gather*}
w^{-1}{\cal L}=\frac{1}{2}D_{\tau}x^{i}\,D_{\tau}x_{i}-\frac{1}{4}\sum_{i\neq k}\left(\big\{x^{i}, x^k\big\}_{W_2}\right)^2.
\end{gather*}
The corresponding bosonic Hamiltonian is
\begin{gather}
H_2=\frac{1}{2}\int_{W_2}{\rm d}^2\sigma\,w(\sigma)^{-1}\,p_i^2(\sigma)
+\frac{1}{4}\sum_{i\neq k}\int_{W_2}{\rm d}^2\sigma\,w(\sigma)\left(\big\{x^{i}(\sigma),x^{k}(\sigma)\big\}_{W_2}\right)^2.
\label{zzd}
\end{gather}
Furthermore there is a constraint whose bosonic part reads
\begin{gather*}
\varphi=\left\{w^{-1}p^{i}, x_{i}\right\}_{W_2}\simeq 0.
\end{gather*}
The above constraint generates the Lie algebra of globally-def\/ined, area-preserving reparametrisations of $W_2$. This latter algebra can be expressed in terms of the Lie brackets (\ref{zy}). For this purpose it suf\/f\/ices to f\/ix a complete orthonormal basis of functions $b_A(\sigma_1, \sigma_2)$, $A=0,1, \ldots$, on~$W_2$. As such they satisfy
\begin{gather}
\int_{W_2}{\rm d}^2\sigma\,w(\sigma)\,b^{A}(\sigma)b_B(\sigma)=\delta^{A}_B
\label{zzf}
\end{gather}
as well as
\begin{gather}
\sum_Ab^{A}(\sigma)b_A(\sigma')=w(\sigma)^{-1}\delta^2(\sigma-\sigma').
\label{zzg}
\end{gather}
Indices are raised and lowered according to
\begin{gather}
b^{A}(\sigma)=(b_A(\sigma))^*=\eta^{AB}\,b_B(\sigma).
\label{zzh}
\end{gather}
Then any function $f=f(\sigma)$ on $W_2$ can be expanded as
\begin{gather}
f(\sigma)=\sum_Af^{A}b_A(\sigma),
\label{zzi}
\end{gather}
and one can write
\begin{gather}
\left\{b_A(\sigma),b_B(\sigma)\right\}_{W_2}=g_{AB}^{\;\;\;C}\,b_C(\sigma)=g_{ABC}\,b^C(\sigma),
\label{zzj}
\end{gather}
where
\begin{gather}
g_{ABC}=\int_{W_2}{\rm d}^2\sigma\,w(\sigma)\,b_A(\sigma)\left\{b_B(\sigma),b_C(\sigma)\right\}_{W_2}.
\label{zzk}
\end{gather}
The $g_{ABC}$ are the structure constants of the (inf\/inite-dimensional) Lie algebra of area-preserving dif\/feomorphisms.

\subsection{Quantum membranes}\label{zzl}

The quantum theory corresponding to the classical Hamiltonian function (\ref{zzd}) requires regulari\-sation \cite{NICOLAI, HOPPE}. This is done by introducing a cutof\/f $\Lambda$ on the inf\/inite range covered by the indices $A,B,C,\ldots$ in equations~(\ref{zzf})--(\ref{zzk}). The inf\/inite number of degrees of freedom present in the classical theory are recovered in the limit $\Lambda\to\infty$. Any f\/inite value of $\Lambda $ corresponds to a truncation of the inf\/inite-dimensional Lie group $G_{W_2}$ of area-preserving reparametrisations of~$W_2$. Truncating means replacing $G_{W_2}$ with some f\/inite-dimensional Lie group $G_{W_2}(\Lambda)$ in such a way that, if the latter has structure constants $\omega_{ABC}(\Lambda)$, then
\begin{gather}
\lim_{\Lambda\to\infty}\omega_{ABC}(\Lambda)=g_{ABC},
\label{zzm}
\end{gather}
where the $g_{ABC}$ are as in equation~(\ref{zzk}). This regularisation leads to a supersymmetric matrix model with a Hamiltonian whose bosonic piece is given by
\begin{gather}
H_2={\rm tr}\,\left(\frac{1}{2}P^{i}P_i+\frac{1}{4}\sum_{i\neq k}[X^{i},X^k]^2\right),
\label{zzn}
\end{gather}
the coordinates $X_i$ and momenta $P_i$ taking values in the Lie algebra of $G_{W_2}(\Lambda)$. In particular, the potential term $\sum\limits_{i\neq k}[X^{i},X^k]^2$ involves the Lie brackets on the Lie algebra of $G_{W_2}(\Lambda)$, which appear as the quantum analogues of the Lie brackets (\ref{zy}) on $W_2$.  Integration over the worldvolume is replaced by tracing over the Lie algebra of $G_{W_2}(\Lambda)$.

It is not guaranteed that a regularised theory satisfying equation~(\ref{zzm}) will exist. The existence of a quantum theory satisfying equation~(\ref{zzm}) depends on the geometry of the membrane. When~$W_2$ is a Riemann sphere it has been proved (see \cite{HOPPE} and references therein) that one can take $G_{W_2}(\Lambda)$ equal to $SU(n)$, with $\Lambda=n^2-1$, in order to obtain a good quantum truncation of the inf\/inite-dimensional group of area-preserving dif\/feomorphisms. Then a convenient complete system of orthonormal functions $b_A(\sigma)$ is given by the usual spherical harmonics, the density function in equation~(\ref{zw}) being $w(\theta, \varphi)=\sin\theta/4\pi$.  Equation~ (\ref{zzm}) was generalised from the sphere to the case when $W_2$ is a torus in~\cite{FAIRLIE}. Equation~(\ref{zzm}) also holds, in the sense of convergence of the structure constants, for an arbitrary compact, quantisable K\"ahler manifold~${\cal M}$ (see below) in any even real dimension $2k$ \cite{BORDE}. In this latter case the $g_{ABC}$ are the structure constants of the Lie algebra of divergence-free dif\/feomorphisms; such reparametrisations preserve symplectic volume on ${\cal M}$, even if the latter (for $k>1$) lacks an interpretation as a membrane worldvolume. However, this property will turn out to be useful for $p$-branes when $p>1$.

{}For the benef\/it of the reader let us recall the def\/inition of a quantisable K\"ahler manifold~${\cal M}$ used above.  Let ${\cal M}$ be a complex manifold and $\omega$ a K\"ahler 2-form on it. Assume that ${\cal M}$ possesses a triple $(L,h,\nabla)$, where $L$ is a holomorphic line bundle on ${\cal M}$, with $h$ a Hermitian metric on this bundle and $\nabla$ a connection on $L$. This connection will also be assumed compatible with the metric $h$ and compatible with the complex structure. If the curvature 2-form $F$ of the connection $\nabla$ on $L$ is such that $F=\omega/{\rm i}\hbar$, then the manifold ${\cal M}$ is called quantisable.

\section[$p$-branes and multibrackets]{$\boldsymbol{p}$-branes and multibrackets}\label{zzo}

\subsection[Classical $p$-branes]{Classical $\boldsymbol{p}$-branes}\label{zzp}

A relativistic $p$-brane embedded within $\mathbb{R}^d$ has a light-cone Hamiltonian whose bosonic piece reads \cite{HOPPE}
\begin{gather}
H_{p}=\frac{1}{2}\int_{W_p}{\rm d}^p\sigma\,w(\sigma)^{-1}p_i^2(\sigma) +\frac{1}{4}\int_{W_p}{\rm d}^p\sigma\,w(\sigma)\,g(\sigma).
\label{zzq}
\end{gather}
Above, the embedding functions $x^{i}=x^{i}(\sigma_1,\ldots, \sigma_p, \tau)$ have the $p_i$ as their conjugate momenta.
The $\sigma_i$, $i=1, \ldots, p$, are spacelike coordinates and $\tau$ is a timelike coordinate parametrising the $p+1$ dimensions of the $p$-brane as the latter evolves in time. We denote by $W_p$ the $p$-brane worldvolume at a f\/ixed value of $\tau$. The function $w(\sigma)$ is a density on $W_p$, and $g(\sigma)$ stands for the determinant of the induced metric
\begin{gather*}
g_{rs}=\partial_r x^{i}\partial_s x_i, \qquad r,s=1, \ldots, p.
\end{gather*}
Def\/ine, for $p$ functions $f_i=f_i(\sigma)$ on $W_p$, the brackets
\begin{gather}
\{f_1,\ldots, f_p\}_{W_p}:= w^{-1}\mbox{alt}\,\left(\partial f_1\cdots\partial f_p\right).
\label{zzs}
\end{gather}
Then one has \cite{HOPPE}
\begin{gather}
g=\sum_{i_1<\dots<i_p}\{x_{i_1},\ldots, x_{i_p}\}_{W_p}\{x^{i_1},\ldots,x^{i_p}\}_{W_p}.
\label{zzt}
\end{gather}
Substituting (\ref{zzt}) into (\ref{zzq}) we obtain the $p$-brane analogue of the bosonic Hamiltonian for the membrane, equation~(\ref{zzd}). Picking a complete, orthonormal set of functions $b_A(\sigma)$ on $W_p$, completeness ensures that their brackets can be expanded as
\begin{gather*}
\left\{b_{A_1}(\sigma), \ldots, b_{A_p}(\sigma)\right\}_{W_p}=g_{A_1\ldots A_p C}\,b^C(\sigma),
\end{gather*}
where
\begin{gather}
g_{A_1\ldots A_p C}=\int_{W_p}{\rm d}^p\sigma\,w(\sigma)\,b_{A_1}(\sigma)\,\left\{b_{A_2}(\sigma), \ldots, b_{A_{p}}(\sigma), b_C(\sigma)\right\}_{W_p}.
\label{zzv}
\end{gather}
The above generalise equations~(\ref{zzj}), (\ref{zzk}). As in the case of the membrane, the $g_{A_1\ldots A_p C}$ are completely antisymmetric and $\sigma$-independent.

\subsection[Quantum $p$-branes]{Quantum $\boldsymbol{p}$-branes}\label{zzz}

We claim that the classical system of Section~\ref{zzp} can be quantised as follows. Let $W_p$ be compact, quantisable and K\"ahler. In particular this implies that $p$ is even, $p=2k$. We measure volumes using the $k$-th exterior power of the symplectic form. We normalise the volume of $W_p$ to unity, as in (\ref{zw}). In this way we can have the cutof\/f $\Lambda$ in equation~(\ref{lawmaker}) below dimensionless. Let $G_{W_p}$ denote the inf\/inite-dimensional Lie group of globally-def\/ined, volume-preserving dif\/feomorphisms of $W_p$.  Then $SU(n)$, for $n\to\infty$, is a quantisation of $G_{W_p}$. Moreover, upon quantisation we would like to replace the classical Nambu brackets~(\ref{zzs}) with some analogous expression containing the Lie-algebra multibrackets of Section~\ref{za}. The Lie algebra $su(n)$ contains the cohomology cocycle $\omega_{2j+1}$, and  then
\begin{gather*}
g=\sum_{i_1<\dots<i_{2j}}[X_{i_1},\ldots, X_{i_{2j}}]_{\omega_{2j+1}}[X^{i_1},\ldots,X^{i_{2j}}]_{\omega_{2j+1}}
\end{gather*}
is the quantisation of equation~(\ref{zzt}).  Moreover, integrals over the worldvolume $W_{p}$ are replaced by traces over a certain irreducible representation of $su(n)$. The quantum Hamiltonian operator~is
\begin{gather*}
H_p={\rm tr}\left(\frac{1}{2}P^{i}P_i+\frac{1}{4}\sum_{i_1<\dots<i_{2j}}[X_{i_1},\ldots, X_{i_{2j}}]_{\omega_{2j+1}}[X^{i_1},\ldots,X^{i_{2j}}]_{\omega_{2j+1}}\right),
\end{gather*}
the coordinates $X_i$ and momenta $P_i$ taking values in some irreducible representation of $su(n)$.

To establish the claim made above it suf\/f\/ices to prove that the following analogue of equation~(\ref{zzm}) holds for the cocycles of order higher than 3:
\begin{gather}
\lim_{\Lambda\to\infty}\omega_{A_1\ldots A_{2j}}^{\qquad\quad C}(\Lambda)= g_{A_1\ldots A_{2j}}^{\qquad\quad C}.
\label{pp}
\end{gather}
Above, the right-hand side is as in equation~(\ref{zzv}); the left-hand side contains the components of the $su(n)$ cocycle $\omega_{2j+1}$. The cutof\/f parameter $\Lambda$ in (\ref{pp}) is the dimension of $su(n)$:
\begin{gather}
\Lambda=n^2-1.
\label{lawmaker}
\end{gather}
In order to prove equation~(\ref{pp}) we use the decompositions  (\ref{zr}) and (\ref{zf}) as applied to the left and right-hand sides of (\ref{pp}), respectively. Assume that the worldvolume $W_p$  satisf\/ies the assumptions stated at the end of Section~\ref{zzl}, namely:  $W_p$ is compact, quantisable and K\"ahler. The left-hand side of (\ref{pp}) decomposes into an antisymmetrised sum of products of twofold Lie brackets on~$su(n)$. The right-hand side of (\ref{pp}) also decomposes into an antisymmetrised sum of products of twofold Poisson brackets. Pick any triple of indices $A$, $B$, $C$ within those two decompositions, and denote the corresponding structure constants by $\omega_{ABC}$ and $g_{ABC}$, respectively.  By equation~(\ref{zzm}), the structure constants $\omega_{ABC}$ of this $su(n)$ provide a quantisation of the corresponding twofold Poisson brackets with structure constants $g_{ABC}$. This allows one to apply (\ref{zzm}) to each and every triple of indices arising within those decompositions. We conclude that, under the assumptions made regarding the worldvolume $W_p$,  equation~(\ref{pp}) holds true. For large enough $n$, the cohomology ring of $su(n)$ always contains a cocycle $\omega_{2j+1}$ such that $2j=2k=p$, where $p$ is the even dimension of $W_p$. Thus the cocycle required  to quantise (\ref{zzs}) is~$\omega_{p+1}$. The algebra $su(n)$ with $n\geq 2$ provides a quantisation of the membrane; for the 4-brane we need $n\geq 3$, for the 6-brane we need $n\geq 4$, and so on.

The careful reader will have noticed the following minor point.  Indices on the right-hand side of (\ref{pp}) are uppercase, running over a complete set of functions on $W_p$. The left-hand side of (\ref{pp}) contains the components of the $su(n)$ cocycle $\omega_{2j+1}$. As such its indices should be lowercase, i.e., Lie-algebra valued. However  the latter have also been written in uppercase. The reason for this is the same as in equation~(\ref{zzm}), whose left-hand side should, in principle, also be lowercase. A vielbein can be introduced on $W_p$ in order to exchange uppercase indices with lowercase indices. Once a vielbein is introduced and applied to the lowest cocycle $\omega_3$ so its indices will be uppercase, the same follows for all higher cocycles $\omega_{2j+1}$.

\section{Discussion}\label{ddkk}

\subsection{Examples}

As a f\/irst example\footnote{We thank the referee for suggesting analysing these two examples.}, let us compare our results with those of the f\/irst reference in \cite{HOPPE}, as applied to the case when the worldvolume $W_p$ is a $p$-dimensional torus, with $p>2$ and even (for example, the product of 4 circles, $S^1\times S^1\times S^1\times S^1$). This manifold qualif\/ies as compact K\"ahler quantisable. As in \cite{HOPPE}, our approach does not quantise by deformation, but rather by compacti\-f\/ication of the relevant inf\/inite-dimensional Lie group of volume-preserving dif\/feomorphisms. Such a compacti\-f\/i\-cation requires truncating the inf\/inite dimension of the Lie group to a f\/inite value, since inf\/inite-dimensional manifolds cannot be compact. Alternatively, one could term our approach as {\it compactification by truncation} of the inf\/inite number of dimensions of the Lie group. However there is no deformation involved in this procedure. In this sense our approach, although perspectively dif\/ferent,  completely agrees with that of \cite{HOPPE}, where a theorem proving the impossibility of deforming the Lie group when the torus has dimension greater than 2 is cleverly circumvented.

Another archetypal example is the case of complex projective space $\mathbb{CP}^N$, also compact K\"ahler quantisable. When $N=1$ this reverts to the case of the Riemann sphere studied in \cite{NICOLAI}. Since this space is $SU(N+1)/SU(N)\times U(1)$, the possibility of approximating its inf\/inite-dimensional Lie group of volume-preserving dif\/feomorphisms by $su(n)$ for $n\to\infty$ transformations is intuitively evident. The manifold $\mathbb{CP}^N$ has a simple (co)homology ring, given by $H_{2j}(\mathbb{CP}^N)=\mathbb{Z}$ for $0\leq j\leq N$ and zero otherwise. This means that, (co)homologically, complex projective space in $N$ complex dimensions can be obtained inductively: starting from a point ($N=0$), we attach a complex 1-dimensional cell at inf\/inity to obtain $\mathbb{CP}^1$; to the latter  we attach a complex 2-dimensional cell at inf\/inity to obtain $\mathbb{CP}^2$, and so forth. This simple inductive structure is ref\/lected in the way one quantises branes whose f\/ixed-time worldvolume equals $\mathbb{CP}^N$. Namely: when $N=1$, the lowest Lie-algebra cocycle $\omega_3$ (necessary to quantise $\mathbb{CP}^1$) is present in $su(n)$, for all $n\geq 2$; when $N=2$, the next lowest Lie-algebra cocycle $\omega_5$ (necessary to quantise $\mathbb{CP}^2$) is present in $su(n)$ for all $n\geq 3$; when $N=3$, the next lowest Lie-algebra cocycle $\omega_7$ (necessary to quantise $\mathbb{CP}^3$) is present in $su(n)$ for all $n\geq 4$; and so forth.

\subsection{Conclusions}

Equation~(\ref{pp}) proves that the Lie-algebra cocycles provide a quantisation of the classical Nambu brackets: the classical limit is recovered when $\Lambda\to\infty$, while any f\/inite value of $\Lambda$ provides a~discrete approximation to the classical, continuous case. As in equation~(\ref{zzm}), this is precisely what quantisation means: the replacement of a continuous bracket operation (on an {\it infinite} number of functions on the worldvolume) with a discrete bracket operation (on a {\it finite} number of generators of a certain Lie algebra).

There are some fundamental similarities between the corresponding quantisations (as summarised by equations~(\ref{zzm}) and (\ref{pp}), respectively). In either case the corresponding limit holds when a specif\/ic choice is made for the manifold representing the worldvolume at f\/ixed time. As indicated above, this approach to quantisation might well be called {\it quantisation by compacti\-fication}: we replace the inf\/inite-dimensional Lie group of volume-preserving dif\/feomorphisms with some compact approximation (or truncation) thereof. Let us explain this point in more detail. No inf\/inite-dimensional manifold can be compact. On the other hand, the Peter--Weyl theorem~\cite{LIE} states that any compact Lie group must necessarily be isomorphic to some subgroup of the unitary group $U(N)$, for some value of $N$~-- in particular, it must be a matrix group. Loosely speaking, {\it quantisation}  is some kind of {\it representation of physical quantities by means of matrices}. Hence our previous statement concerning quantisation as a compactif\/ication (of the group of volume-preserving dif\/feomorphisms, in our case) makes perfectly good sense.

A power of ${\rm i}\hbar$ must multiply all time derivatives in the quantum theory, as in equation~(\ref{zk}). This power must appear on the left-hand side of expressions relating time evolution to multibrackets, so no such factor need appear in our quantisation prescription (\ref{pp}). For quantum-mechanical consistency, where the classical limit is def\/ined as that in which $\hbar\to 0$, Planck's constant must be understood as being inversely proportional to the cutof\/f parameter $\Lambda$ in equation~(\ref{lawmaker}).

 Finally it should be mentioned that (supersymmetric) matrix models such as (\ref{zzn}) arise in M-theory \cite{WITTEN, BFSS}, where the limit $n\to\infty$ is also taken. Related topics have been analysed in~\cite{GALAVIZ} and \cite{NOI},
respectively.

\subsection*{Acknowledgements}
It is a great pleasure for J.M.I.  to thank Max-Planck-Institut f\"ur Gravitationsphysik, Albert-Einstein-Institut, Golm (Germany) for hospitality during the preparation of this work. The authors thank Generalitat Valenciana (Spain) for f\/inancial support.

\pdfbookmark[1]{References}{ref}
\LastPageEnding

\end{document}